\documentclass[aps,prl,twocolumn,groupedaddress,showpacs]{revtex4}

\usepackage{graphicx}

\begin{document}

\title{Surface states and the charge of a dust particle in a plasma}
%\title{Surface model for the charge of a dust particle in a plasma}

\author{F. X. Bronold$^1$, H. Fehske$^1$,  H. Kersten$^2$, and H. Deutsch$^{1}$}
\address{$^1$Institut f\"ur Physik, Ernst-Moritz-Arndt-Universit\"at Greifswald, D-17489
Greifswald, Germany}
\affiliation{$^2$ Institut f\"ur Experimentelle und Angewandte Physik,
Christian-Albrechts-Universit\"at zu Kiel, D-24098
Kiel, Germany}

\date{\today}

%\begin{abstract}
%We investigate electron and ion surface states of a negatively
%charged dust particle in a plasma and propose an effective surface
%model to calculate the charge of the particle form the balance of
%charging and desorption fluxes. Our approach shows that electrons
%desorb from polarization-induced surface states, a few Angstroms 
%away from the grain surface, whereas ions escape from shallow states
%corresponding to the largest stable classical orbits in the
%presence of charge-exchange scattering. Using an orbital-motion
%limited charging flux for electrons and the grain temperature as an 
%adjustable parameter, we obtain excellent agreement with experimental 
%data.
%\end{abstract}

\begin{abstract}
We investigate electron and ion surface states of a negatively
charged dust particle in a gas discharge and identify the charge of 
the particle with the electron surface density bound in the 
polarization-induced short-range part of the particle 
potential. On that scale, ions do not affect the charge. They are 
trapped in the shallow states of the Coulomb tail of the potential 
and act only as screening charges. 
Using orbital-motion limited electron charging fluxes and the 
particle temperature as an adjustable parameter, we obtain excellent 
agreement with experimental data.
\end{abstract}

\maketitle

\noindent
{\it Motivation.---} 
The calculation of the charge of a macroscopic object in an ionized gas
is one of the most fundamental problems of plasma physics. It occurs 
in space-bound plasmas, where the charge of satellites is
of interest~\cite{Whipple81}, in astrophysical plasmas, where one wants
to know the charge of interstellar grains~\cite{Horanyi96}, and
in laboratory gas discharges, where charged dust particles are either 
contaminants, which need to be controlled, or constituents, 
whose collective properties are the subject of study~\cite{Ishihara07,FIK05}. 

For laboratory plasmas, the particle charge has been measured
in a number of experiments~\cite{KRZ05,SV03,TAA00,TLA00,WHR95}. 
Throughout it is thereby assumed that the particle's surface 
potential, and hence its charge, is the one which balances at 
the grain surface the total electron with the total ion charging 
flux: $j_e=j_i$. In almost all cases, however, the charges obtained
from this condition, which is equivalent to forcing the net charge of the
particle to be quasi-stationary, are too high. Usually, the approximations
for the fluxes, mostly orbital motion limited (OML)~\cite{KA03,AAA00,BR59}, 
are blamed for the disagreement and various modifications of the OML theory
have been proposed. Although this leads sometimes to reasonable quantitative
results~\cite{SV03,KRZ05}, we suspect on fundamental grounds that irrespective 
of the fluxes $j_e=j_i$ is not the condition which fixes the charge (or potential)
of the particle.

The condition $j_e=j_i$ is part of the Boltzmann-Poisson description of the 
plasma-particle interaction. Its natural length scale is thus the length on 
which the Coulomb potential varies. There are however microscopic processes near 
the surface of the particle, most notable sticking and desorption of electrons, 
which affect the charge but take place on a much shorter scale. Once these 
processes are incorporated it is clear that the charge of the particle 
is not determined by the quasi-stationarity of the net charge but by the 
individual quasi-stationarity of the electron and ion densities bound to 
the particle. This condition implies the former but not vice versa. It is 
thus more restrictive and leads to lower charges.

In this Letter we describe a surface model, which accounts for 
plasma- and surface-induced processes, and calculate the charge of the 
particle, and its partial screening due to trapped ions, without relying 
on the condition $j_e=j_i$. Instead, we force the electron and ion 
densities bound to the particle to be quasi-stationary by balancing, 
individually on effective surfaces, electron and ion charging fluxes 
with electron and ion desorption fluxes.

{\it Surface states.---}
We start with an investigation of the bound states 
in the {\it static} interaction potential of an electron (ion) and 
a dust particle with radius $R$, dielectric constant $\epsilon$, and charge $-eZ_p$. 
The potential contains a polarization-induced part, arising from the electric boundary 
condition at the grain surface, and a Coulomb tail due to the particle's charge~\cite{Boettcher52}. 
Defining $\xi=(\epsilon-1)/2(\epsilon+1)Z_p$, measuring distances from the grain 
surface in units of $R$ and energies in units of $\bar{U}=Z_pe^2/R$, the interaction 
energy at $x=r/R-1>x_b$, where $x_b$ is a cut-off below which the grain
surface is not perfect anymore, reads 
\begin{eqnarray}
V_{e,i}(x)&=&\pm\frac{1}{1+x}-\frac{\xi}{x(1+x)^2(2+x)}
\nonumber\\
&\approx&\left\{\begin{array}{ll}
1-\xi/2x & \mbox{electron}\\
-1/(1+x) & \mbox{ion}~.
\end{array}\right.
\label{V(x)}
\end{eqnarray}

The second line approximates the relevant parts of the interaction energy
very well and permits an analytical calculation of surface states. A gas 
discharge usually contains enough electrons which can overcome the 
particle's Coulomb barrier $\bar{U}\sim o(eV)$. These are the electrons 
which may get bound in the polarization-induced short-range part 
of the potential, well described by the approximate expression. Ions, 
on the other hand, having a finite radius 
$r^{size}_i/R=x_i^{size}\approx 10^{-4}$,
cannot explore the potential at these distances. The long-range 
Coulomb tail is most relevant to them, which is again well described by the 
approximate expression. 

In order to determine bound states from the Schr\"odinger equations 
corresponding to $V_{e,i}(x)$ we have 
to specify boundary conditions. Clearly, the wavefunctions $u_{e,i}(x)$ have 
to vanish for $x\rightarrow\infty$, irrespective of the potentials. The 
boundary condition at $x_b$, in contrast, depends on the potential for 
$x\le x_b$, that is, on the surface barrier. 
For our purpose, it is sufficient to take the simplest barrier model: 
$V_{e,i}(x\le x_b)=\infty$ with $x_b=0$ for electrons and $x_b=x_i^{size}$
for ions. The wave functions vanish then also at $x_b$ and the surface 
states are basically rescaled hydrogen-type wavefunctions.
\begin{figure}[t]
\includegraphics[width=0.5\linewidth]{Fig1new.eps}\includegraphics[width=0.5\linewidth]{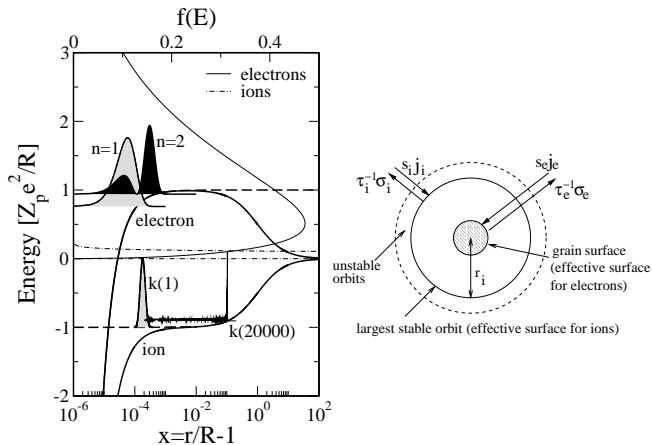}
\caption{\label{Potential}
Left panel: Potential energy for an electron (ion) in the
field of a MF particle ($R=4.7~\mu m$, $Z=6800$)~\cite{KDK04} 
and s-type probability distributions 
shifted to the binding energy and maxima normalized to one.
Dashed lines are the potentials used in the 
Schr\"odinger equations and thin lines are the bulk energy 
distribution functions for the hosting discharge.
%Note, the finite ion radius $r_i^{size} \sim \AA$ 
%forces the ion wavefunctions to vanish at $x\approx 10^{-4}$. 
Right panel: Illustration of the surface model to be 
discussed in the main text. At quasi-stationarity, surface 
charges $\sigma_{e,i}$ bound at $r_e\approx R$ and 
$r_i\approx (2\pi\sigma_{\rm cx}n_g)^{-1}$, respectively, 
balance charging fluxes $s_{e,i}j_{e,i}$ with
desorption fluxes $\tau_{e,i}^{-1}\sigma_{e,i}$.
}
\end{figure}

The left panel of Fig.~\ref{Potential} shows s-type electron and ion probability densities
$|u_{e,i}(x)|^2$ (our reasoning does not depend on the angular momentum)
for a melamine-formaldehyde (MF) particle 
($R=4.7~\mu m$, $\epsilon=8$, and $Z_p=6800$) in a helium discharge with 
plasma density $n_e\approx n_i=0.62\times 10^9~cm^{-3}$, ion temperature
$k_BT_i=0.04~eV$, and electron temperature $k_BT_e=2.2~eV$~\cite{KDK04}.
The Rydberg series 
of electron surface states is only a few Angstroms away from the grain boundary. 
At these distances, the spatial variation of $V_e(x)$ is comparable to the 
de-Broglie wavelength of electrons approaching the particle: 
$\lambda_e^{dB}/R\sim |V_e/V_e'|\sim 10^{-4}$. Hence, the trapping 
of electrons at the surface of the particle is a 
quantum-mechanical effect not included in the classical 
description of the plasma-particle interaction. For ions, on the other 
hand, the lowest surface states, which carry quantum-mechanical features,
are unimportant. Being cold and heavy, ions will be bound in 
a continuum of states below the ion ionization threshold which consists 
essentially of classical trapped orbits, as can be seen from the $|u_i(x)|^2$ 
for the $k(20000)$ state. That ions behave classically is not unexpected. 
Their de-Broglie wavelength is much smaller then the scale on which the 
potential varies:  
$\lambda_i^{dB}/R\sim 10^{-5}\ll |V_i/V_i'|\sim 1$ for $x \gtrsim 10^{-3}$.
%Thus, the interaction between ions and the particle can be 
%described classically.

{\it Model.---} We now use the properties of the surface states to construct 
a model for the charge of the particle. Within the sheath of
the particle, the density of free electrons (ions) is much smaller 
than the density of bound electrons (ions). In that region, the quasi-stationary 
charge (in units of $-e$) is thus approximately 
%given by
\begin{eqnarray}
Z(x)&=&4\pi R^3\int_{x_b}^x dx' \big(1+x'\big)^2 \bigg[n^b_e(x')-n^b_i(x')
\bigg]
\label{Start}
\end{eqnarray}
with $x<\lambda^D_i=\sqrt{kT_i/e n_i}$, the ion Debye length, which 
we take as an upper cut-off, and $n^b_{e,i}$ the density of bound 
electrons and ions. The results presented above suggest to express 
the density of bound electrons by an electron
surface density:  $n^b_e(x)\approx\sigma_e\delta(x-x_e)/R$ with 
$x_e\sim x_b\sim 0$ and $\sigma_e$ the quasi-stationary solution of
\begin{eqnarray}
d\sigma_e/dt=s_e j_e-\tau_e^{-1}\sigma_e~,
\label{rate}
\end{eqnarray}
where $j_e$ is the electron charging flux from the plasma and $s_e$ and 
$\tau_e$ are, respectively, the electron sticking coefficient and electron 
desorption time due to {\it inelastic} collisions between electrons and the
particle (see right panel in Fig.~\ref{Potential}). We will argue below that 
once the particle has collected some negative charge, not necessarily the 
quasi-stationary one, there is a critical ion orbit at 
$x_i\sim 1-10 \gg x_e$ which prevents ions from hitting the particle 
surface. Thus, the particle charge is simply
\begin{eqnarray}
Z_p\equiv Z(x_e<x<x_i)=4\pi R^2 (s\tau)_{e} j_{e}~.
\label{Zp}
\end{eqnarray}

For an electron to get stuck at (to desorb from) a surface it has 
to loose (gain) energy at (from) the surface. Since electrons with
rather low and rather high energies are, respectively, reflected by 
the Coulomb and surface barrier of the particle, sticking (desorption)
primarily affects electrons at energies slightly above $\bar{U}$. 
Assuming this group of electrons to be in quasi-equilibrium 
with the surface electrons after overcoming the Coulomb barrier,
absolute reaction rate theory~\cite{KG86} allows us to estimate
\begin{eqnarray}
(s\tau)_e\approx
\frac{h}{k_BT_p}\exp\bigg[\frac{E_e^d}{k_BT_p}\bigg]~,
\label{stau}
\end{eqnarray}
where $h$ is Planck's constant, $T_p$ is the particle temperature,
and $E_e^d$ is the negative of the binding energy of the surface
state from which desorption most likely occurs. This 
phenomenological equation relates a combination of kinetic 
coefficients, which individually depend on the dynamic interaction, 
to an energy which can be deduced from the static interaction alone.
To go beyond Eq.~(\ref{stau}) necessitates a quantum-kinetic 
treatment of the inelastic electron-particle interaction.
\begin{figure}[t]
\includegraphics[width=\linewidth]{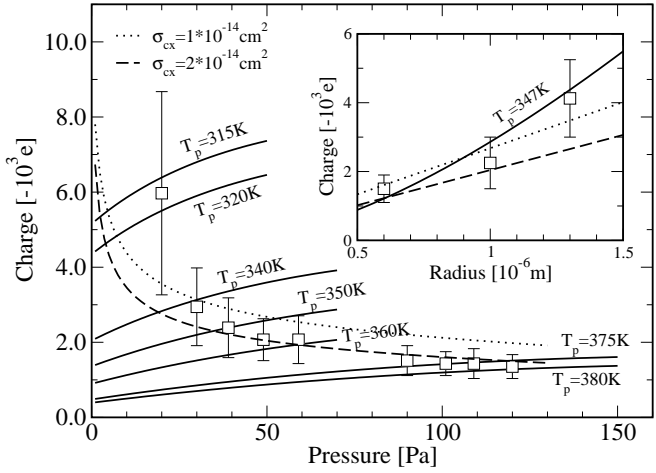}
\caption{\label{Khrapak}
Pressure dependence of the charge of a MF particle with $R=1~\mu m$
in the neon discharge of Ref.~\cite{KRZ05} (squares). Solid lines 
denote the (isothermal) charges deduced from the surface model 
whereas dotted and dashed lines are the charges obtained from
$j_e^{\rm OML}=j_i^{\rm OML}+j_i^{\rm CX}$. The inset shows the 
radius dependence of the charge for $p=50~Pa$.
}
\end{figure}

Equation~(\ref{Zp}) is a self-consistency equation for $Z_p$. 
More explicitly, combined with Eq.~(\ref{stau}), and using 
the OML electron charging flux, which is a reasonable approximation 
because, on the plasma scale, electrons are repelled from the 
particle, it reads 
\begin{eqnarray}
Z_p=4\pi R^2\frac{h}{k_BT_p}e^{E_e^d/k_BT_p}j_e^{\rm OML}(Z_p) 
\label{Zfinal}
\end{eqnarray}
with $j_e^{\rm OML}=n_e\sqrt{k_BT_e/2\pi m_e}\exp[-Z_pe^2/Rk_BT_e]$. 
Thus, $Z_p$ depends on the radius $R$,
the plasma parameters $n_e$ and $T_e$, and the surface 
parameters $T_p$ and $E_e^d$. 

{\it Results.---} To estimate $E_e^d$ we imagine an 
electron with energy just above $\bar{U}$ approaching the 
grain. By necessity, it comes very close to the surface (see left 
panel in Fig.~\ref{Potential}). For any realistic surface barrier, 
the wavefunction will therefore leak into the grain and the electron
will strongly couple to the excitations of the grain which 
provide the thermal reservoir encoded in $T_p$. Hence, the electron
will quickly relax to the lowest surface state. The $n=1$ state 
for the infinitely high barrier is an approximation to that state.
Thus, $E_e^d\approx R_0(\epsilon-1)^2/16(\epsilon+1)^2$,
where $R_0$ is the Rydberg energy.
$T_p$ cannot be determined as simply.
It depends on the heating and cooling fluxes 
to-and-fro the grain and thus on additional surface 
parameters~\cite{SKD00}. We use $T_p$ therefore as an adjustable parameter. 
To reproduce, for instance, with Eq.~(\ref{Zfinal}) the charge of 
the particle in Fig.~\ref{Potential}, 
$T_p=395~K$ implying $E_e^d\approx 0.51~eV$ and 
$(s\tau)_e\approx 0.4\times 10^{-6}~s$.
\begin{figure}[t]
\includegraphics[width=\linewidth]{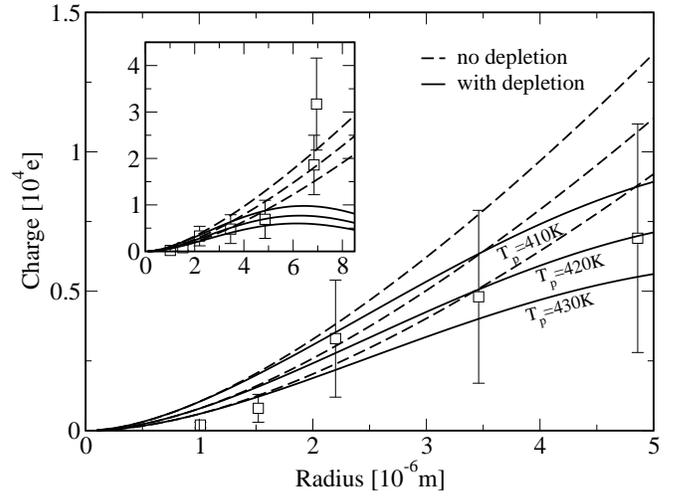}
\caption{\label{Tomme}
Radius dependence of the charge of a MF particle in the sheath of
an argon discharge at $p=6.67~Pa$~\cite{TAA00} (squares). Solid
(dashed) lines give the charges deduced from the surface
model when the depletion of $n_e$ in the confining 
sheath is (is not) included in the OML electron charging flux.
}
\end{figure}

In Fig.~\ref{Khrapak} we analyze within our approach the pressure dependence 
of the charge of a MF particle with $R=1~\mu m$ in the bulk of the neon discharge
of Ref.~\cite{KRZ05}. Since the plasma parameters entering Eq.~(\ref{Zfinal})
are known~\cite{KRZ05}, $T_p$ is again the only free parameter. 
Fixing $T_p$ at a particular value gives the isothermal particle charges 
$Z_p(T_p)$. From $Z_p(T_p)=Z_{exp}$ follows then the $T_p$ required 
to reproduce the data. The predicted increase of $T_p$ with pressure is 
realistic. Indeed, assuming $T_p\sim T_g$, with $T_g$ the gas 
temperature, $T_p$ is in accordance with what one would expect from 
the pressure dependence of $T_g$ in noble gases~\cite{SKD00}.
For comparison we also plot the particle charges deduced from
$j_e^{\rm OML}=j_i^{\rm OML}+j_i^{\rm CX}$ with 
$j_i^{\rm OML}=n_i\sqrt{k_BT_i/2\pi m_i}[1+Z_pe^2/R k_BT_i]$
the OML ion charging flux and 
$j_i^{\rm CX}=n_i(0.1\lambda^D_i/l_{\rm cx})
\sqrt{k_BT_i/2\pi m_i}(Z_pe^2/R k_BT_i)^2$ 
the ion flux due to charge-exchange (CX), where 
$l_{\rm cx}=(\sigma_{\rm cx} n_g)^{-1}$ is the scattering length,
$\sigma_{\rm cx}$ is the cross section, and $n_g=p/k_BT_g$ is the gas 
density~\cite{KRZ05}. Although for $\sigma_{\rm cx}=2\times 10^{-14}~cm^2$
(which may be unrealistically large~\cite{HES82}) 
the agreement with the data is good, the radius dependence of $Z_p$ 
at fixed pressure shown in the inset indicates that something must be 
wrong with the flux balance criterion. The data clearly appear to be closer
to the non-linear $R-$dependence obtained from the surface model than 
to the linear one resulting from $j_e^{\rm OML}=j_i^{\rm OML}+j_i^{\rm CX}$.

Figure~\ref{Tomme}, showing the $R-$dependence of $Z_p$ for MF particles
confined in the sheath of an argon discharge~\cite{TAA00}, provides 
additional support for our model. To approximately account for the 
fact that particles with different radius experience different 
plasma environments, we included the depletion of $n_e$ in 
the sheath by replacing $n_e$ in $j_{e}^{\rm OML}$ by
$n_e\exp[e\Phi(z_{\rm eq}(R))/k_BT_e]$ with $\Phi(z)$ the sheath potential
and $z_{\rm eq}(R)$ the equilibrium position of the particle with 
radius $R$~\cite{TAA00}. When the grains are not
too deep in the sheath ($R<5~\mu m$), we find excellent agreement
with the data for $T_p=420~K$. Our approach fails, however, for 
$R>5~\mu m$ (see inset). We attribute this to the ad-hoc description of $j_e$
%which in fact should be self-consistently calculated taking, among others,
%sub-thermal electrons from the electrode into account~\cite{SV03}.
which may not capture the total electron charging flux close to the
electrode.

Equation~(\ref{Zfinal}) depends on the assumption that once the 
particle is negatively charged ions are trapped far away from 
the grain surface. Indeed, a recent study based on the Boltzmann-Poisson 
equations has shown that charge-exchange collisions lead to a local 
pile-up of ions in the sheath of the particle~\cite{LGS03,LGG01}. 
We come to the same conclusion from the surface physics point of view.
Similar to an electron, an ion gets bound to the grain only when 
it looses energy. Because of the long-range attractive ion-grain
interaction, the ion will be initially bound far away from the
grain surface (see left panel in Fig.~\ref{Potential}).
The coupling to the excitations
of the grain is thus negligible and only inelastic processes due to 
the plasma are able to induce transitions to lower bound states.
Since the interaction is classical, inelastic collisions, for instance, 
charge-exchange between ions and atoms, act like a random force.
Energy relaxation can be thus envisaged as a de-stabilization of orbits 
whose spatial extension is comparable to or larger then the scattering 
length. Smaller orbits are unaffected because the collision probability during 
one revolution is vanishingly small. For a circular orbit, a rough estimate 
for the critical radius is $r_i=R(1+x_i)=(2\pi\sigma_{\rm cx} n_g)^{-1}$  
which leads to $x_i\sim 5\gg x_e \sim 0$ when we use the parameters 
of the helium discharge of Fig.~\ref{Potential} and 
$\sigma_{\rm cx}=0.32\times 10^{-14}~cm^2$~\cite{M56}. Thus, 
there is a relaxation bottleneck at $x_i$ and ions are trapped deep 
in the sheath of the particle.

To determine the partial screening due to trapped ions we model
the ion density $n_i^b$ accumulating in the vicinity of the critical
orbit by a surface density $\sigma_i$ which balances at $x_i$ the ion 
charging flux with the ion desorption flux (see right panel in 
Fig.~\ref{Potential}). Mathematically, this gives rise to a rate equation 
similar to~(\ref{rate}) but now for the ions. At quasi-stationarity, the ion 
surface density is thus $\sigma_i=(s\tau)_i j_i$. Although Eq.~(\ref{stau})
assumes excitations of the grain to be responsible for 
sticking and desorption we expect a similar expression (with $E_e^d$, 
$T_p$ replaced by $E_i^d$, $T_g$) to control the density 
of trapped ions. From Eq.~(\ref{Start}) we then obtain 
$Z(x_i<x<\lambda^D_i)=Z_p-Z_i$ with 
\begin{eqnarray}
Z_i=4\pi R^2(1+x_i)^2
\frac{h}{k_BT_g}e^{E_i^d(Z_p)/k_BT_g}j^B_i~
\label{Zion}
\end{eqnarray}
the number of trapped ions. Since the critical orbit is near 
the particle-sheath-plasma boundary, it is fed by 
the Bohm ion flux $j^B_i=0.6n_i\sqrt{k_BT_e/m_i}$. The
ion desorption energy is the negative of the binding energy 
of the critical orbit,
$E_i^d(Z_p)=-V_i(x_i)\bar{U}(Z_p)=4\pi\sigma_{\rm cx}a_B n_g Z_pR_0$,
and depends strongly on $Z_p$ and $x_i$. 
For the situation shown in Fig.~\ref{Potential} we obtain
$E_i^d\approx 0.37~eV$ and $(s\tau)_i\approx 0.6\times 10^{-8}~s$ 
when we use $T_g=T_p=395~K$, the particle temperature which 
reproduces $Z_p\approx 6800$. The ion screening charge is then
$Z_{i}\approx 148\ll Z_p$ which is the order of magnitude 
expected from molecular dynamics simulations~\cite{CK94}. Thus, 
even when the particle charge is defined by $Z(x_i<x<\lambda^D_i)$ 
it is basically given by $Z_p$. 

{\it Summary.---} We constructed a surface model to 
calculate the charge (partial screening) of a particle
in a plasma by balancing, on an effective surface, the electron (ion) 
charging with the electron (ion) desorption flux. The number of electrons
bound in the polarization potential determines the charge of the 
particle. Using the grain temperature as an adjustable parameter 
we obtained far better agreement with measurements, in particular, with 
respect to the radius dependence of the charge, then approaches 
based on balancing at the grain surface the total charging fluxes 
which we argue is the wrong condition. It neglects the microscopic
processes determining the charge of the particle: 
sticking and desorption of electrons at the grain surface.

\begin{acknowledgments}
Support from the SFB-TR~24 is greatly acknowledged.
F.~X.~B. is funded by MV 0770/461.01.
%by the state
%Mecklenburg-Vorpommern.
\end{acknowledgments}


\begin{thebibliography}{21}

\expandafter\ifx\csname natexlab\endcsname\relax\def\natexlab#1{#1}\fi
\expandafter\ifx\csname bibnamefont\endcsname\relax
  \def\bibnamefont#1{#1}\fi
\expandafter\ifx\csname bibfnamefont\endcsname\relax
  \def\bibfnamefont#1{#1}\fi
\expandafter\ifx\csname citenamefont\endcsname\relax
  \def\citenamefont#1{#1}\fi
\expandafter\ifx\csname url\endcsname\relax
  \def\url#1{\texttt{#1}}\fi
\expandafter\ifx\csname urlprefix\endcsname\relax\def\urlprefix{URL }\fi
\providecommand{\bibinfo}[2]{#2}
\providecommand{\eprint}[2][]{\url{#2}}

\bibitem[{\citenamefont{Whipple}(1981)}]{Whipple81}
\bibinfo{author}{\bibfnamefont{E.~C.} \bibnamefont{Whipple}},
  \bibinfo{journal}{Rep. Prog. Phys.} \textbf{\bibinfo{volume}{44}},
  \bibinfo{pages}{1197} (\bibinfo{year}{1981}).

\bibitem[{\citenamefont{Hor\'anyi}(1996)}]{Horanyi96}
\bibinfo{author}{\bibfnamefont{M.}~\bibnamefont{Hor\'anyi}},
  \bibinfo{journal}{Annu. Rev. Astron. Astrophys.}
  \textbf{\bibinfo{volume}{34}}, \bibinfo{pages}{383} (\bibinfo{year}{1996}).

\bibitem[{\citenamefont{Ishihara}(2007)}]{Ishihara07}
\bibinfo{author}{\bibfnamefont{O.}~\bibnamefont{Ishihara}},
  \bibinfo{journal}{J. Phys. D: Appl. Phys.} \textbf{\bibinfo{volume}{40}},
  \bibinfo{pages}{R121} (\bibinfo{year}{2007}).

\bibitem[{\citenamefont{Fortov et~al.}(2005)\citenamefont{Fortov, Ivlev,
  Khrapak, Khrapak, and Morfill}}]{FIK05}
\bibinfo{author}{\bibfnamefont{V.~E.} \bibnamefont{Fortov}},
  \bibinfo{author}{\bibfnamefont{A.~V.} \bibnamefont{Ivlev}},
  \bibinfo{author}{\bibfnamefont{S.~A.} \bibnamefont{Khrapak}},
  \bibinfo{author}{\bibfnamefont{A.~G.} \bibnamefont{Khrapak}},
  \bibnamefont{and} \bibinfo{author}{\bibfnamefont{G.~E.}
  \bibnamefont{Morfill}}, \bibinfo{journal}{Phys. Rep.}
  \textbf{\bibinfo{volume}{421}}, \bibinfo{pages}{1} (\bibinfo{year}{2005}).

\bibitem[{\citenamefont{Khrapak et~al.}(2005)\citenamefont{Khrapak, Ratynskaia,
  Zobnin, Usachev, Yaroshenko, Thoma, Kretschmer, Hoefner, Morfill, Petrov
  et~al.}}]{KRZ05}
\bibinfo{author}{\bibfnamefont{S.~A.} \bibnamefont{Khrapak}},
  \bibinfo{author}{\bibfnamefont{S.~V.} \bibnamefont{Ratynskaia}},
  \bibinfo{author}{\bibfnamefont{A.~V.} \bibnamefont{Zobnin}},
  \bibinfo{author}{\bibfnamefont{A.~D.} \bibnamefont{Usachev}},
  \bibinfo{author}{\bibfnamefont{V.~V.} \bibnamefont{Yaroshenko}},
  \bibinfo{author}{\bibfnamefont{M.~H.} \bibnamefont{Thoma}},
  \bibinfo{author}{\bibfnamefont{M.}~\bibnamefont{Kretschmer}},
  \bibinfo{author}{\bibfnamefont{H.}~\bibnamefont{Hoefner}},
  \bibinfo{author}{\bibfnamefont{G.~E.} \bibnamefont{Morfill}},
  \bibinfo{author}{\bibfnamefont{O.~F.} \bibnamefont{Petrov}},
  \bibnamefont{et~al.}, \bibinfo{journal}{Phys. Rev. E}
  \textbf{\bibinfo{volume}{72}}, \bibinfo{pages}{016406}
  (\bibinfo{year}{2005}).

\bibitem[{\citenamefont{Samarian and Vladimirov}(2003)}]{SV03}
\bibinfo{author}{\bibfnamefont{A.~A.} \bibnamefont{Samarian}} \bibnamefont{and}
  \bibinfo{author}{\bibfnamefont{S.~V.} \bibnamefont{Vladimirov}},
  \bibinfo{journal}{Phys. Rev. E} \textbf{\bibinfo{volume}{67}},
  \bibinfo{pages}{066404} (\bibinfo{year}{2003}).

\bibitem[{\citenamefont{Tomme et~al.}(2000{\natexlab{a}})\citenamefont{Tomme,
  Annaratone, and Allen}}]{TAA00}
\bibinfo{author}{\bibfnamefont{E.~B.} \bibnamefont{Tomme}},
  \bibinfo{author}{\bibfnamefont{B.~M.} \bibnamefont{Annaratone}},
  \bibnamefont{and} \bibinfo{author}{\bibfnamefont{J.~E.} \bibnamefont{Allen}},
  \bibinfo{journal}{Plasma Sources Sci. Technol.} \textbf{\bibinfo{volume}{9}},
  \bibinfo{pages}{87} (\bibinfo{year}{2000}{\natexlab{a}}).

\bibitem[{\citenamefont{Tomme et~al.}(2000{\natexlab{b}})\citenamefont{Tomme,
  Law, Annaratone, and Allen}}]{TLA00}
\bibinfo{author}{\bibfnamefont{E.~B.} \bibnamefont{Tomme}},
  \bibinfo{author}{\bibfnamefont{D.~A.} \bibnamefont{Law}},
  \bibinfo{author}{\bibfnamefont{B.~M.} \bibnamefont{Annaratone}},
  \bibnamefont{and} \bibinfo{author}{\bibfnamefont{J.~E.} \bibnamefont{Allen}},
  \bibinfo{journal}{Phys. Rev. Lett.} \textbf{\bibinfo{volume}{85}},
  \bibinfo{pages}{2518} (\bibinfo{year}{2000}{\natexlab{b}}).

\bibitem[{\citenamefont{Walch et~al.}(1995)\citenamefont{Walch, Hor\'anyi, and
  Robertson}}]{WHR95}
\bibinfo{author}{\bibfnamefont{B.}~\bibnamefont{Walch}},
  \bibinfo{author}{\bibfnamefont{M.}~\bibnamefont{Hor\'anyi}},
  \bibnamefont{and}
  \bibinfo{author}{\bibfnamefont{S.}~\bibnamefont{Robertson}},
  \bibinfo{journal}{Phys. Rev. Lett.} \textbf{\bibinfo{volume}{75}},
  \bibinfo{pages}{838} (\bibinfo{year}{1995}).

\bibitem[{\citenamefont{Kennedy and Allen}(2003)}]{KA03}
\bibinfo{author}{\bibfnamefont{R.~V.} \bibnamefont{Kennedy}} \bibnamefont{and}
  \bibinfo{author}{\bibfnamefont{J.~E.} \bibnamefont{Allen}},
  \bibinfo{journal}{J. Plasma Phys.} \textbf{\bibinfo{volume}{69}},
  \bibinfo{pages}{485} (\bibinfo{year}{2003}).

\bibitem[{\citenamefont{Allen et~al.}(2000)\citenamefont{Allen, Annaratone, and
  de~Angelis}}]{AAA00}
\bibinfo{author}{\bibfnamefont{J.~E.} \bibnamefont{Allen}},
  \bibinfo{author}{\bibfnamefont{B.~M.} \bibnamefont{Annaratone}},
  \bibnamefont{and}
  \bibinfo{author}{\bibfnamefont{U.}~\bibnamefont{de~Angelis}},
  \bibinfo{journal}{J. Plasma Phys.} \textbf{\bibinfo{volume}{63}},
  \bibinfo{pages}{299} (\bibinfo{year}{2000}).

\bibitem[{\citenamefont{Bernstein and Rabinowitz}(1959)}]{BR59}
\bibinfo{author}{\bibfnamefont{I.~B.} \bibnamefont{Bernstein}}
  \bibnamefont{and} \bibinfo{author}{\bibfnamefont{I.~N.}
  \bibnamefont{Rabinowitz}}, \bibinfo{journal}{Phys. Fluids}
  \textbf{\bibinfo{volume}{2}}, \bibinfo{pages}{112} (\bibinfo{year}{1959}).

\bibitem[{\citenamefont{Boettcher}(1952)}]{Boettcher52}
\bibinfo{author}{\bibfnamefont{C.~J.~F.} \bibnamefont{Boettcher}},
  \emph{\bibinfo{title}{Theory of electric polarization}}
  (\bibinfo{publisher}{Elsevier Publishing Company},
  \bibinfo{address}{Amsterdam}, \bibinfo{year}{1952}).

\bibitem[{\citenamefont{Kersten et~al.}(2004)\citenamefont{Kersten, Deutsch,
  and Kroesen}}]{KDK04}
\bibinfo{author}{\bibfnamefont{H.}~\bibnamefont{Kersten}},
  \bibinfo{author}{\bibfnamefont{H.}~\bibnamefont{Deutsch}}, \bibnamefont{and}
  \bibinfo{author}{\bibfnamefont{G.~M.~W.} \bibnamefont{Kroesen}},
  \bibinfo{journal}{Int. J. Mass Spectrometry} \textbf{\bibinfo{volume}{233}},
  \bibinfo{pages}{51} (\bibinfo{year}{2004}).

\bibitem[{\citenamefont{Kreuzer and Gortel}(1986)}]{KG86}
\bibinfo{author}{\bibfnamefont{H.~J.} \bibnamefont{Kreuzer}} \bibnamefont{and}
  \bibinfo{author}{\bibfnamefont{Z.~W.} \bibnamefont{Gortel}},
  \emph{\bibinfo{title}{Physisorption Kinetics}} (\bibinfo{publisher}{Springer
  Verlag}, \bibinfo{address}{Berlin}, \bibinfo{year}{1986}), \bibinfo{pages}{pp. 13}.

\bibitem[{\citenamefont{Swinkels et~al.}(2000)\citenamefont{Swinkels, Kersten,
  Deutsch, and Kroesen}}]{SKD00}
\bibinfo{author}{\bibfnamefont{G.~H. P.~M.} \bibnamefont{Swinkels}},
  \bibinfo{author}{\bibfnamefont{H.}~\bibnamefont{Kersten}},
  \bibinfo{author}{\bibfnamefont{H.}~\bibnamefont{Deutsch}}, \bibnamefont{and}
  \bibinfo{author}{\bibfnamefont{G.~M.~W.} \bibnamefont{Kroesen}},
  \bibinfo{journal}{J. Appl. Phys.} \textbf{\bibinfo{volume}{88}},
  \bibinfo{pages}{1747} (\bibinfo{year}{2000}).

\bibitem[{\citenamefont{Hegerberg et~al.}(1982)\citenamefont{Hegerberg, Elford,
  and Skullerud}}]{HES82}
\bibinfo{author}{\bibfnamefont{R.}~\bibnamefont{Hegerberg}},
  \bibinfo{author}{\bibfnamefont{M.~T.} \bibnamefont{Elford}},
  \bibnamefont{and} \bibinfo{author}{\bibfnamefont{H.~R.}
  \bibnamefont{Skullerud}}, \bibinfo{journal}{J. Phys. B}
  \textbf{\bibinfo{volume}{15}}, \bibinfo{pages}{797} (\bibinfo{year}{1982}).

\bibitem[{\citenamefont{Lampe et~al.}(2003)\citenamefont{Lampe, Goswami,
  Sternovsky, Robertson, Gavrishchaka, Ganguli, and Joyce}}]{LGS03}
\bibinfo{author}{\bibfnamefont{M.}~\bibnamefont{Lampe}},
  \bibinfo{author}{\bibfnamefont{R.}~\bibnamefont{Goswami}},
  \bibinfo{author}{\bibfnamefont{Z.}~\bibnamefont{Sternovsky}},
  \bibinfo{author}{\bibfnamefont{S.}~\bibnamefont{Robertson}},
  \bibinfo{author}{\bibfnamefont{V.}~\bibnamefont{Gavrishchaka}},
  \bibinfo{author}{\bibfnamefont{G.}~\bibnamefont{Ganguli}}, \bibnamefont{and}
  \bibinfo{author}{\bibfnamefont{G.}~\bibnamefont{Joyce}},
  \bibinfo{journal}{Phys. Plasma} \textbf{\bibinfo{volume}{10}},
  \bibinfo{pages}{1500} (\bibinfo{year}{2003}).

\bibitem[{\citenamefont{Lampe et~al.}(2001)\citenamefont{Lampe, Gavrishchaka,
  Ganguli, and Joyce}}]{LGG01}
\bibinfo{author}{\bibfnamefont{M.}~\bibnamefont{Lampe}},
  \bibinfo{author}{\bibfnamefont{V.}~\bibnamefont{Gavrishchaka}},
  \bibinfo{author}{\bibfnamefont{G.}~\bibnamefont{Ganguli}}, \bibnamefont{and}
  \bibinfo{author}{\bibfnamefont{G.}~\bibnamefont{Joyce}},
  \bibinfo{journal}{Phys. Rev. Lett.} \textbf{\bibinfo{volume}{86}},
  \bibinfo{pages}{5278} (\bibinfo{year}{2001}).

\bibitem[{\citenamefont{Moiseiwitsch}(1956)}]{M56}
\bibinfo{author}{\bibfnamefont{B.~L.} \bibnamefont{Moiseiwitsch}},
  \bibinfo{journal}{Proc. Phys. Soc. London A} \textbf{\bibinfo{volume}{69}},
  \bibinfo{pages}{653} (\bibinfo{year}{1956}).

\bibitem[{\citenamefont{Choi and Kushner}(1994)}]{CK94}
\bibinfo{author}{\bibfnamefont{S.~J.} \bibnamefont{Choi}} \bibnamefont{and}
  \bibinfo{author}{\bibfnamefont{M.~J.} \bibnamefont{Kushner}},
  \bibinfo{journal}{IEEE Trans. Plasma Science} \textbf{\bibinfo{volume}{22}},
  \bibinfo{pages}{138} (\bibinfo{year}{1994}).

\end{thebibliography}
\end{document}